# Measuring optical transmission matrices by wavefront shaping

Jonghee Yoon[1,2], KyeoReh Lee[1,2], Jongchan Park[1], and YongKeun Park[1,*]

[1]*Department of Physics, KAIST, Daejeon 305-701, South Korea*
[2]*Contributed equally to this work*
[*]*yk.park@kaist.ac.kr*

**Abstract:** We introduce a simple but practical method to measure the optical transmission matrix (TM) of complex media. The optical TM of a complex medium is obtained by modulating the wavefront of a beam impinging on the complex medium and imaging the transmitted full-field speckle intensity patterns. Using the retrieved TM, we demonstrate the generation and linear combination of multiple foci on demand through the complex medium. This method will be used as a versatile tool for coherence control of waves through turbid media.

## 1. Introduction

Light transport through complex media is a fundamental optical phenomenon relevant to various research fields and applications from light localization [1], quantum secure authentication [2], and random lasers [3] to imaging through highly turbid media such as biological tissue [4]. Recently, research has revealed the potential of coherent control of multiple scattered light [5] that exploits the linear relationship between incidents and transmitted light fields through a scattering medium, which is described by a transmission matrix (TM).

One class of approaches to control multiple light scattering involves wavefront shaping techniques [6] that manipulate the wavefront of a beam impinging on a scattering medium in such a way as to control the transmitted light field. Employing a spatial light modulator (SLM), the optical wavefront before the scattering medium is optimized by monitoring the transmitted light using various feed-back algorithms [6-8]. The advantage of the wavefront shaping technique lies in the simple optical system, which has been utilized in recent works demonstrating a high degree control of scattering light fields in space [6, 9], time [10], wavelength [11], polarization [12], imaging [13], transmission energy [14], at sub-wavelength scales [15], and photo-acoustically guided light [16]. However, unfortunately, such methods only access one output mode at a time; as many optimization processes should be performed as the number of optical modes of interest. In spite of the simplicity of the method, this limitation prevents practical applications of wavefront shaping techniques.

Measuring the TM of a turbid medium is an elegant technique for coherent control of multiple scattered light fields because light transmission through a scattering medium is completely described by the TM. Recently, TM approaches have shown potential in focusing [17], delivering images [18], controlling transmitted energy [19], subwavelength imaging [20, 21], multispectral control [22], and acoustically modulated light [21]. While TM approaches provide full control of light fields, technical limitations in the existing methods prevent wider utilizations of the TM approaches. *Popoff et al.* first demonstrated the optical TM measurement [17] using a common-path interferometric setup equipped with a SLM. However, this approach only a portion of SLM pixels was used to modulate modes of incident fields, while the remaining SLM pixels had to be kept static for a reference beam, which limited the number of measurable modes in the TM measurements. In addition, the Hadamard basis can only address *N* channels for $2^{2N}$ SLM pixels used, which would be only a portal of accessible TM information. Alternatively, an interferometer equipped with a rotating galvano-meter based mirror [23] can directly measure the complex amplitudes of transmitted light, from which the TM can be obtained. However, these approaches require complicated optical setups such as Mach-Zehnder interferometry and the setups usually suffer from stabilization issues due to phase noise.

Here we introduce a simple but powerful method to directly measure the optical TM of turbid media exploiting wavefront shaping techniques. Our approach is based on the principle that information about optimized wavefronts at the incident side that result in the generation of optical foci at every single point at the output side is exactly the same information as that of the TM of the system. By modulating the wavefront of incident beams using a SLM and simultaneously collecting consequent two-dimensional intensity maps of transmitted speckle fields, a series of optimized wavefront information to generate individual foci at the transmitted side can be obtained at the same time. From this information, the optical TM of a turbid medium can be directly reconstructed. Our method uses only the simple optical setup used in conventional wavefront shaping techniques, but it directly provides full TM measurements that had previously been obtained only by using complicated interferometric approaches.

The paper is organized as follows. In Sec. II, we define the problem and present the detailed principle of the algorithm. The experimental setup and the sample preparation are described in Sec. III. In Sec. IV, we present the measurement of a TM of a highly scattering layer. Furthermore, we demonstrate the generation of multiple foci through the turbid layer on demand using the measured TM information. The approach for directly measuring an optical TM with a simple setup for wavefront shaping can be readily used in various direct applications in complex optics.

## 2. Principles

### 2.1 Problem definition

The concept of an optical system discussed in this paper is illustrated in Fig. 1. Defining each SLM pixel on an orthonormal basis, $|x_1\rangle, |x_2\rangle, \cdots, |x_N\rangle$, a light field transmitted through an optical system can be expressed as,

$$|\psi\rangle = \sum_{n=1}^{N} \mathbf{t} |x_n\rangle, \qquad (1)$$

where $\mathbf{t}$ represents a TM of any linear optical system including turbid media, and $N$ is the total number of SLM pixels. In cases of turbid media, $|\psi\rangle$ is a speckle light field resulting from multiple light scattering events. When an arrayed image sensor or a camera is used for imaging the transmitted light field, the camera pixels are another set of an orthonormal basis of interest, $|y_1\rangle, |y_2\rangle, \cdots, |y_M\rangle$, where $M$ is the total number of camera pixels. Then, the optical field assigned to the $m$-th camera pixel $|y_m\rangle$ can be expressed as a linear combination of the input basis

$$\langle y_m | \psi \rangle = \sum_{n=1}^{N} \langle y_m | \mathbf{t} | x_n \rangle. \qquad (2)$$

Thus, measuring the TM of an optical imaging system $\langle y_m | \mathbf{t} | x_n \rangle$ or $t_{mn}$ for all $(m,n)$ pairs provides the full information about light transport in the system.

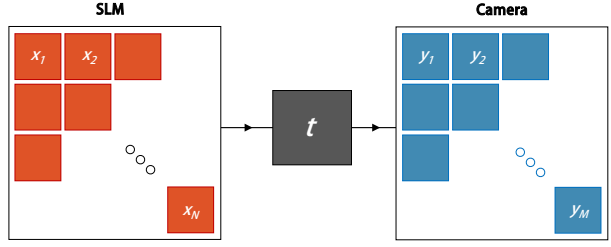

Fig. 1. Conceptual diagram of the optical system. The optical field is transferred from a SLM to a camera through a linear optical system (***t***).

*2.2 Procedure to measure a transmission matrix*

The proposed algorithm measures the TM of a turbid layer using a wavefront shaping method that is expanded from the parallel wavefront optimization method [24]. Whereas the previous parallel wavefront optimization method finds the optimized wavefront of an illumination beam that generates a focus through random scattering media, our approach utilizes the full-field speckle intensity maps in order to simultaneously obtain the optimized phase maps that give rise to focusing at various positions through scattering media.

In order to fully exploit pixels in the SLM, we divided the SLM pixels into two groups, as shown in Figs. 2(a)–(b). One group modulates the impinging wavefront ('signal beam'), while the other part remains fixed in order to provide controlled reference beams. We applied the phase information to the each group in a checkerboard pattern to ensure the reference field has light intensity comparable to the signal field over the field of view in a camera.

For convenience, let the signal and reference groups in the SLM plane consist of pixels $|x_1\rangle, |x_2\rangle, \cdots, |x_G\rangle$ and $|x_{G+1}\rangle, |x_{G+2}\rangle, \cdots, |x_N\rangle$, respectively. As proposed in the parallel wavefront optimization method [24], we assigned distinct modulating frequencies to different SLM pixels in the single group. The zero frequency is assigned to SLM pixels in the reference group. This procedure enables the parallel modulation of SLM pixels, which results in fast and robust wavefront shaping. Then, the optical field transmitting a scattering medium can be described as a sum of the reference $|R\rangle$ and sample $|S\rangle$ fields,

$$|\psi\rangle = |R\rangle + |S\rangle = \sum_{n=G+1}^{N} \mathbf{t}|x_n\rangle + \sum_{n=1}^{G} \mathbf{t}|x_n\rangle e^{i\omega_n t}, \qquad (3)$$

where *N* and *G* are the total number of SLM pixels in the reference and sample groups, respectively. Inserting Eq. (3) into Eq. (2), the measured intensity map at the detector plane can be expressed as

$$|\langle y_m|\psi\rangle|^2 = |\langle y_m|R\rangle|^2 + \sum_{p,q=1}^{G} t_{mp}^* t_{mq} e^{i(\omega_q - \omega_p)t}$$
$$+ \langle y_m|R\rangle^* \sum_{p=1}^{G} t_{mp} e^{i\omega_p t} + \langle y_m|R\rangle \sum_{p=1}^{G} t_{mp}^* e^{-i\omega_p t}. \qquad (4)$$

Because the third term in Eq. (4) is the information of interest to construct a TM, we need to decouple the third term from the others. To achieve the goal, we measured a series of full-field intensity maps at the detector plane (total number of the series is 4G) with $\omega_p = \frac{G+p}{4G}\omega_0$, where $\omega_0$ is the frame rate of the detector. Then, the harmonic frequency $\omega_q - \omega_p$ should be found in the range $\left[-\frac{1}{2}\omega_0, \frac{1}{2}\omega_0\right]$, and this range does not overlap with $\omega_p$. Thus, using Fourier transformation, the term $\langle y_m|R\rangle^* t_{mp}$ can be selectively extracted from the measured series of intensity maps. Exchanging the roles of groups, all SLM pixels can be utilized for modulating fields. Thus, the minimum number of intensity measurements is 4*N*.

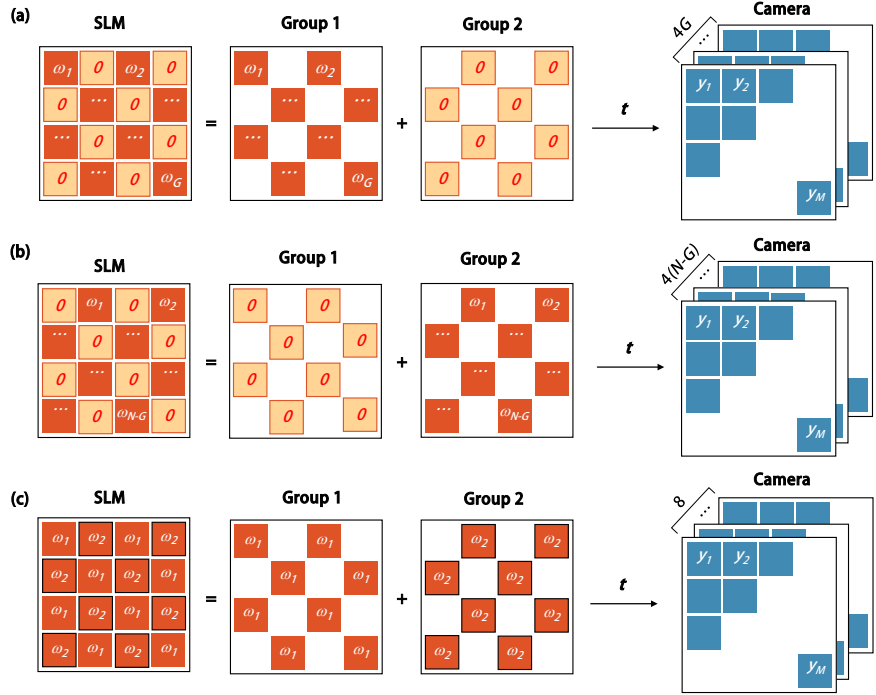

Fig. 2. Schematics for each step of the TM measurement algorithm. (a) Group1 measurement step. Individual pixels of Group 1 is modulated, while Group 2 stays still to provide a reference field. Required acquisitions are 4*G*. (b) Group2 measurement step. Identical with (a) but the role of groups are exchanged. Required acquisitions are 4(*N-G*). (c) Reference phase matching step. Each group rather than each pixel is modulated individually. Required acquisitions are 8.

*2.3 Reference phase matching*

In order to construct an exact TM from the retrieved optimized phase patterns, it is crucial to access the reference fields. Two reference fields $|R_1\rangle$ and $|R_2\rangle$ obtained using the parallel wavefront shaping method described in the above section resulted from the reference groups in the SLM pixels $|x_1\rangle, |x_2\rangle, \cdots, |x_G\rangle$ and $|x_{G+1}\rangle, |x_{G+2}\rangle, \cdots, |x_N\rangle$, respectively. For simplicity without losing generality, we also considered an ambient reference field $|R_0\rangle$. In an ideal situation, $|R_0\rangle$ should be zero; however, in practical experimental situations, an ambient reference field exists which results from unmodulated light in a SLM, fixed diffraction patterns due to the pixilation in a SLM, multiple reflections in optics, etc. Then, the retrieved optimized phase patterns in Section 2.2 can be expressed as

$$\text{Group 1}: (r_{m0} + r_{m2})^* (t_{m1}, t_{m2}, \cdots, t_{mG})$$
$$\text{Group 2}: (r_{m0} + r_{m1})^* (t_{m(G+1)}, t_{m(G+2)}, \cdots, t_{mN})$$
(5)

where $\langle y_m | R_x \rangle \equiv r_{mx}$ for the simple notation.

For TM construction, it is important compensate for two different reference fields in Eq. (5). For this purpose, we introduce an algorithm for reference phase matching, as illustrated in Fig. 2(c). We modulated all the pixels in each group in the SLM with two characteristic frequencies, $\omega_1$ and $\omega_2$, and measured the corresponding series of speckle intensity patterns,

$$|\psi\rangle = |R_0\rangle + |R_1\rangle e^{i\omega_1 t} + |R_2\rangle e^{i\omega_2 t}, \tag{6}$$

where $|R_0\rangle$ is the ambient reference field. Equation (6) can be considered as a case with $G = 2$ in Eq. (3). Therefore, eight minimum measurements of speckle intensity maps are required to extract the field information about interference terms $r_{m0}^* r_{m1}$, $r_{m0}^* r_{m2}$, and $r_{m2}^* r_{m1}$ from the modulation frequencies $\omega_1$, $\omega_2$, and $\omega_1 - \omega_2$, respectively. Then, the compensation term $C$ can be obtained as

$$\begin{aligned} C &= (r_{m0} + r_{m2})^* (r_{m0} + r_{m1}) = |r_{m0}|^2 + r_{m0}^* r_{m1} + r_{m2}^* r_{m0} + r_{m2}^* r_{m1} \\ &= \frac{(r_{m0}^* r_{m2})^* (r_{m0}^* r_{m1})}{r_{m2}^* r_{m1}} + r_{m0}^* r_{m1} + (r_{m0}^* r_{m2})^* + r_{m2}^* r_{m1}, \end{aligned} \tag{7}$$

or simply $C = r_{m2}^* r_{m1}$ for the ideal case $r_{m0} = 0$. Applying the normalized $C$, the measurements in (5) can be rewritten as

$$\begin{aligned} \text{Group 1} &: \frac{|r_{m0} + r_{m2}|}{|r_{m0} + r_{m2}|}(r_{m0} + r_{m2})^* (t_{m1}, t_{m2}, \cdots, t_{mG}) \\ \text{Group 2} &: \frac{|r_{m0} + r_{m1}|}{|r_{m0} + r_{m2}|}(r_{m0} + r_{m2})^* (t_{m(G+1)}, t_{m(G+2)}, \cdots, t_{mN}). \end{aligned} \tag{9}$$

The shared phase-only term $(r_{m0} + r_{m2})^* / |r_{m0} + r_{m2}|$ can be removed from both groups because it is a global phase constant, which leads to matching of reference phases for both groups as

$$\begin{aligned} \text{Group 1} &: |r_{m0} + r_{m2}|(t_{m1}, t_{m2}, \cdots, t_{mG}) \\ \text{Group 2} &: |r_{m0} + r_{m1}|(t_{m(G+1)}, t_{m(G+2)}, \cdots, t_{mN}). \end{aligned} \tag{10}$$

From the obtained information in (10), the phase part of the TM can be directly constructed. We note that the amplitude of reference fields $|r_{m0} + r_{m2}|$ and $|r_{m0} + r_{m1}|$ can also be measured by alternatively turning off each group from which both amplitude and phase parts of the TM can be constructed.

## 3. Methods

### 3.1. Optical setup

The experimental setup for measuring TMs using the present method is shown in Fig. 3. A diode pumped solid state (DPSS) laser ($\lambda = 532$ nm, 100 mW, Shanghai Dream Lasers Technology, China) is used for an illumination source. The beam from the laser is spatially filtered and expanded using a 4-$f$ telescopic imaging system. The input and output polarization states of the beams to a spatial light modulator (X10468-01, Hamamatsu Photonics Inc., Japan) are maintained to be linearly polarized using two polarizers (LPVISE100-A, Thorlabs Inc., USA). The polarization direction of the incident light to the SLM is aligned to the working direction of the SLM, which modulate the phase of an outgoing wave while maintaining the polarization states. The modulated beam from the SLM is projected onto a surface of a scattering sample via an objective lens (60×, NA = 0.8, CFI 60X, Nikon, Japan) and a 4-$f$ telescopic imaging system. The demagnification factor from the SLM to the sample plane was set as ×600 to ensure that the projected size of SLM pixels is smaller than the diffraction-limited spot, or the minimum size of propagating optical modes. The light transmitting through the scattering sample was collected using an objective lens (60×, NA = 0.8, CFI 60X, Nikon, Japan) and then projected on a CMOS camera (Lt365R, pixel size = 4.54 μm, Lumnera Inc., USA) via a 4-*f* telescopic imaging system.

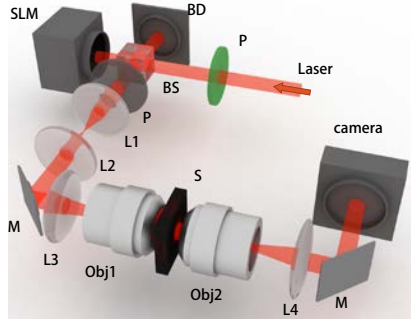

Fig. 3. Experimental setup. Used abbreviations are defined as followings: BD, beam dump; SLM, spatial light modulator; P, polarizer; L#, lens; M, mirror; Obj#, objective lens; S, scattering medium.

*3.2. Scattering sample preparation*

Turbid media used in the study were made of commercial spray paint (Pingo General, Noroo Paint, South Korea), consists of $TiO_2$ nanoparticles with a mean diameter of 200 nm and fixative polymer [15]. The paint is sprayed onto a glass coverslip, and the thickness of a scattering layer was measured as 39.2 μm. The effective refractive index of the scattering layer was calculated as 1.787 using elemental analysis obtained by energy dispersive x-ray spectroscopy.

## 4. Results and discussions

Using the optical setup and the algorithm described in Section 3, we measured the TM of a turbid layer made of spray paint containing $TiO_2$ nanoparticles. The phase part of the reconstructed scattering layer $\angle \mathbf{t} = \sum_{m=1}^{M} \sum_{n=1}^{N} \angle t_{mn} |y_m\rangle \langle x_n|$ is shown in Fig. 4(a). The row and column of the TM are the output and input channels, which correspond to the pixel indexes of the camera and the SLM, respectively. Figure 4(b) shows the reshaped 2-D map of the *m*-th row of the conjugated TM, $\angle(\mathbf{t}^\dagger |y_m\rangle)$, which corresponds to the optimal wavefront information to be applied to the SLM in order to generate an optical focus at the *m*-th camera pixel.

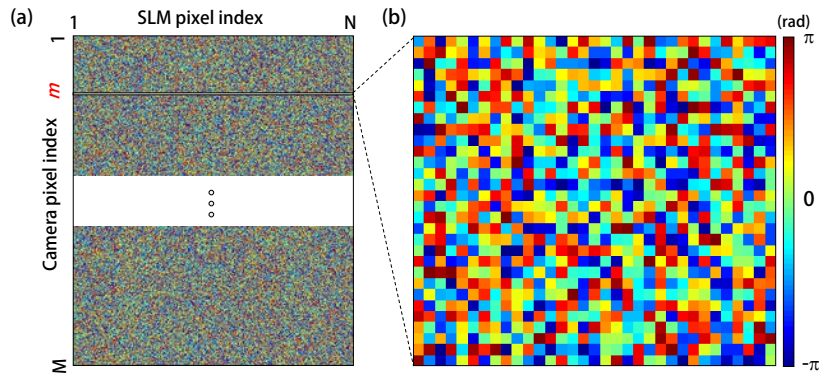

Fig. 4. Retrieved optical TM of a turbid medium. (a) The phase information of the retrieved conjugated TM. Row and column correspond to the pixel indexes of the camera and SLM, respectively. (b) The reshaped 2-D map of the *m*-th row of the conjugated TM, which corresponds to the optimal wavefront information to be applied to the SLM in order to generate an optical focus at the *m*-th camera pixel.

In order to validate the accuracy of the proposed method, we retrieved optimal wavefront maps that gave rise to foci at specific points at the CCD plane [e.g., Fig. 4(b)]. Then we applied these optimal phase maps to the SLM and measured corresponding intensity maps at the CCD plane. Figure 5(a) shows movements of the optimized focus along a diagonal direction beyond the scattering sample by applying a series of columns of the measured TM. A clear focus is formed at the target position. This result clearly demonstrates the accuracy of the measured TM. Because the single TM measurement provides full information about the wavefront of impinging beams to generate a focus at every point on the CCD plane, we can generate an optical focus at an arbitrary point without any further optimization processes. The intensity enhancement factors $\eta$, defined as the ratio of the intensity of generated foci over the averaged intensity in backgrounds, were 198 with 990 SLM segments. This enhancement is approximately 20% of the theoretically expected value ($\eta = 777.76$) [6]. We attribute this difference to the inhomogeneous voltage-phase response functions of SLM pixels, phase digitization due to the limited bit-depth of the SLM (158 digits for $2\pi$ in used wavelength), and the laser intensity fluctuation.

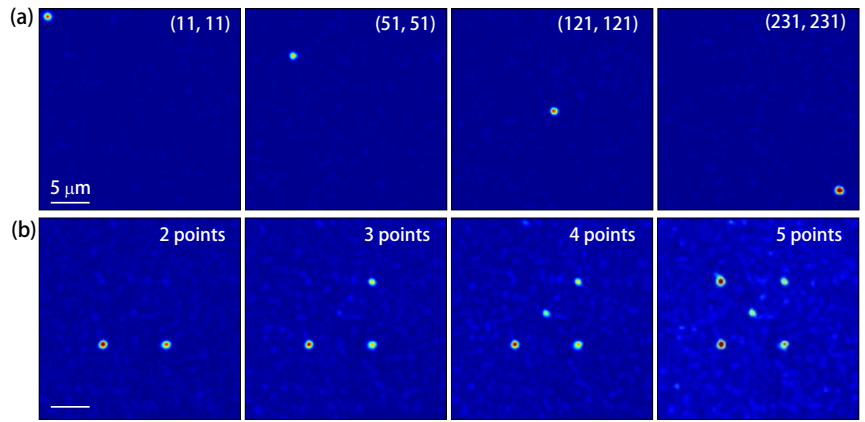

Fig. 5. Focusing through a turbid medium using retrieved TM information. (a) Diagonal focus movements by displaying the corresponding optimal wavefront on the SLM. The numbers above each image indicate the (*x*, *y*) coordinate of target foci. (b) Multiple foci formation with a linear combination of optimal wavefronts.

The total time for measuring a TM with a size of 990 × 36,864 was $1.32 \times 10^3$ sec. In the present method, the time for measuring a TM is comparable to the parallel wavefront shaping technique, which is only able to generate a single point. The measurement speed in the current study is mainly limited by the refresh rate of the SLM. The measurement speed can be can further enhanced by employing fast wavefront modulating devices, including a dynamic mirror device or a deformable mirror [25-27].

To demonstrate the applicability, we demonstrate that the optimized phase map to generate multiple foci at the CCD plane can be directly obtained from the measured TM. The optimal phase map for the impinging beam in order to generate multiple foci on *Y* different positions can be obtained from a linear combination of optimal wavefronts to generate foci on the wanted positions or a sum of the corresponding rows of the TM corresponds to the optimal phase map to generate multiple foci:

$$\left|SLM_{opt}\right\rangle = \angle\left(\sum_{m}^{Y}\frac{1}{\sqrt{Y}}\mathbf{t}^{\dagger}\left|y_{m}\right\rangle\right). \quad (11)$$

By adding multiple rows of the TM and applying the combined phase map to the SLM, the generation of multiple foci ($Y$ = 2, 3, 4, and 5) was obtained as shown in Fig. 5(b). Multiple foci were clearly generated by applying the phase information calculated from the measured TM. Again, the TM information can be fully obtained via one wavefront shaping process; the generation of multiple foci on demand can be readily achieved without performing additional wavefront shaping or optimization processes.

To further exploit the expandability of the proposed method, we also demonstrate the controlled distribution of transmission energy to multiple foci generated using the TM information. The intensities of individual foci through complex media can be controlled by multiplying real-valued weighting factors $\alpha_m$ when obtaining the optimal phase map as

$$\left|SLM_{opt}\right\rangle = \angle\left(\sum_{m}^{Y}\alpha_{m}\mathbf{t}^{\dagger}\left|y_{m}\right\rangle\right). \quad (12)$$

where $\sum_{m}^{X}\alpha_{m}^{2}=1$. Using conventional wavefront shaping methods, one can generate multiple foci in a single optimization process, but it is impossible to control the intensities of individual foci. One may only achieve the goal by performing a series of wavefront shaping processes in conventional approaches [13]. The present approach, however, achieves the TM measurement in one optimization process, and thus one can directly generate multiple foci with individually controlled intensities. Figure 6 shows the results for a case of $Y=2$. The simple addition of weighting factor $\alpha_1$ enables the redistribution of energy to each focus in a controlled manner.

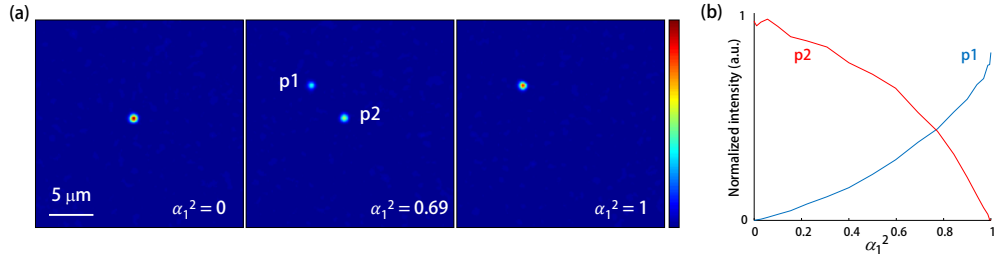

Fig. 6. Controlling the intensities of generated foci. (a) Images of two optimized foci obtained with a different fraction value $\alpha_1^2$. (b) the intensities of two optimized foci as a function of the fraction value.

## 5. Conclusion

In summary, we report a simple, practical, but powerful method to measure a TM by simultaneously multiplexing pixels in a SLM and a detector. By performing one optimization setup for wavefront shaping, a TM of an optical system can be directly retrieved with a simple imaging system. This is the first report on combining both wavefront shaping and TM measurement approaches. In fact, the algorithms and discussions presented in this paper indicate that these two different approaches address the same exact information in light transport through complex media.

Using the proposed method, we measured the TM of a highly scattering medium consisting of $TiO_2$ nanoparticles. Utilizing the measured TM, we demonstrate the generation of an optical focus at different positions and multiple foci with controlled intensities. The proposed approach will be particularly useful in coherent control of multiple light scattering because (1) the proposed method uses only a simple optical setup for wavefront shaping, which is unlike conventional approaches based on interferometric microscopy, (2) the measure speed is very fast; only one wavefront shaping process is enough to generate full TM information, (3) the pixel resolution of a SLM is fully utilized, which enables a high degree of control. Furthermore, we note that digital optical phase conjugation [28-32] can also be addressed with the present method.

Considering the growing importance of light control in complex media, we suggest that the proposed direct method for measuring optical TM will open up new avenues of various studies, including focusing, imaging, and sensing in various applications ranging from non-invasive biomedical optics, addressing light localization regions such as in Anderson localization or Levy walks.

Acknowledgements

This work was supported by National Research Foundation (NRF) of Korea (2012R1A1A1009082, 2014K1A3A1A09063027, 2013R1A1A3011886, 2012-M3C1A1-048860, 2013M3C1A3063046, 2014M3C1A3052537), KAIST, KU-KAIST program, and APCTP program.

**References**

1. D. S. Wiersma, P. Bartolini, A. Lagendijk, and R. Righini, "Localization of light in a disordered medium," **390**(6661), 671-673 (1997).
2. S. A. Goorden, M. Horstmann, A. P. Mosk, B. Škorić, and P. W. Pinkse, "Quantum-secure authentication of a physical unclonable key," **1**(6), 421-424 (2014).
3. H. Cao, Y. Zhao, S. Ho, E. Seelig, Q. Wang, and R. Chang, "Random laser action in semiconductor powder," **82**(11), 2278 (1999).
4. V. V. Tuchin, and V. Tuchin, *Tissue optics: light scattering methods and instruments for medical diagnosis* (SPIE press Bellingham, 2007).
5. A. P. Mosk, A. Lagendijk, G. Lerosey, and M. Fink, "Controlling waves in space and time for imaging and focusing in complex media," **6**(5), 283-292 (2012).
6. I. M. Vellekoop, and A. Mosk, "Focusing coherent light through opaque strongly scattering media," **32**(16), 2309-2311 (2007).
7. I. Vellekoop, and A. Mosk, "Phase control algorithms for focusing light through turbid media," **281**(11), 3071-3080 (2008).
8. D. B. Conkey, A. N. Brown, A. M. Caravaca-Aguirre, and R. Piestun, "Genetic algorithm optimization for focusing through turbid media in noisy environments," **20**(5), 4840-4849 (2012).
9. T. Cizmar, M. Mazilu, and K. Dholakia, "In situ wavefront correction and its application to micromanipulation," **4**(6), 388-394 (2010).
10. J. Aulbach, B. Gjonaj, P. M. Johnson, A. P. Mosk, and A. Lagendijk, "Control of light transmission through opaque scattering media in space and time," **106**(10), 103901 (2011).
11. J. H. Park, C. H. Park, H. Yu, Y. H. Cho, and Y. K. Park, "Active spectral filtering through turbid media," **37**(15), 3261-3263 (2012).
12. J.-H. Park, C. Park, H. Yu, Y.-H. Cho, and Y. Park, "Dynamic active wave plate using random nanoparticles," **20**(15), 17010-17016 (2012).
13. T. Čižmár, and K. Dholakia, "Exploiting multimode waveguides for pure fibre-based imaging," **3**(1027 (2012).

14. I. Vellekoop, and A. Mosk, "Universal optimal transmission of light through disordered materials," **101**(12), 120601 (2008).
15. J.-H. Park, C. Park, H. Yu, J. Park, S. Han, J. Shin, S. H. Ko, K. T. Nam, Y.-H. Cho, and Y. Park, "Subwavelength light focusing using random nanoparticles," **7**(6), 454-458 (2013).
16. P. Lai, L. Wang, J. W. Tay, and L. V. Wang, "Photoacoustically guided wavefront shaping for enhanced optical focusing in scattering media," (2015).
17. S. Popoff, G. Lerosey, R. Carminati, M. Fink, A. Boccara, and S. Gigan, "Measuring the transmission matrix in optics: an approach to the study and control of light propagation in disordered media," **104**(10), 100601 (2010).
18. S. Popoff, G. Lerosey, M. Fink, A. C. Boccara, and S. Gigan, "Image transmission through an opaque material," **1**(81 (2010).
19. M. Kim, Y. Choi, C. Yoon, W. Choi, J. Kim, Q.-H. Park, and W. Choi, "Maximal energy transport through disordered media with the implementation of transmission eigenchannels," **6**(581-585 (2012).
20. C. Park, J.-H. Park, C. Rodriguez, H. Yu, M. Kim, K. Jin, S. Han, J. Shin, S. H. Ko, and K. T. Nam, "Full-Field Subwavelength Imaging Using a Scattering Superlens," **113**(11), 113901 (2014).
21. T. Chaigne, O. Katz, A. C. Boccara, M. Fink, E. Bossy, and S. Gigan, "Controlling light in scattering media non-invasively using the photoacoustic transmission matrix," **8**(1), 58-64 (2014).
22. D. Andreoli, G. Volpe, S. Popoff, O. Katz, S. Gresillon, and S. Gigan, "Deterministic control of broadband light through a multiply scattering medium via the multispectral transmission matrix," (2014).
23. H. Yu, T. R. Hillman, W. Choi, J. O. Lee, M. S. Feld, R. R. Dasari, and Y. Park, "Measuring Large Optical Transmission Matrices of Disordered Media," Phys Rev Lett **111**(15), 153902 (2013).
24. M. Cui, "Parallel wavefront optimization method for focusing light through random scattering media," **36**(6), 870-872 (2011).
25. D. B. Conkey, A. M. Caravaca-Aguirre, and R. Piestun, "High-speed scattering medium characterization with application to focusing light through turbid media," **20**(2), 1733-1740 (2012).
26. J. Jang, J. Lim, H. Yu, H. Choi, J. Ha, J. H. Park, W. Y. Oh, W. Jang, S. Lee, and Y. Park, "Complex wavefront shaping for optimal depth-selective focusing in optical coherence tomography," Opt Express **21**(3), 2890-2902 (2013).
27. H. Yu, J. Jang, J. Lim, J.-H. Park, W. Jang, J.-Y. Kim, and Y. Park, "Depth-enhanced 2-D optical coherence tomography using complex wavefront shaping," **22**(7), 7514-7523 (2014).
28. M. Cui, and C. Yang, "Implementation of a digital optical phase conjugation system and its application to study the robustness of turbidity suppression by phase conjugation," **18**(4), 3444-3455 (2010).
29. T. R. Hillman, T. Yamauchi, W. Choi, R. R. Dasari, M. S. Feld, Y. Park, and Z. Yaqoob, "Digital optical phase conjugation for delivering two-dimensional images through turbid media," **3**((2013).
30. I. M. Vellekoop, M. Cui, and C. Yang, "Digital optical phase conjugation of fluorescence in turbid tissue," **101**(8), 081108-081108-081104 (2012).
31. X. Xu, H. Liu, and L. V. Wang, "Time-reversed ultrasonically encoded optical focusing into scattering media," **5**(3), 154-157 (2011).
32. B. Judkewitz, Y. M. Wang, R. Horstmeyer, A. Mathy, and C. H. Yang, "Speckle-scale focusing in the diffusive regime with time reversal of variance-encoded light (TROVE)," Nat Photonics **7**(4), 300-305 (2013).